\documentclass[12pt]{article}
\usepackage{graphicx, amsmath, amssymb, cite, setspace, color, relsize, bm, bbold, array, hyperref, dsfont, bbold, xcolor, cancel,soul,placeins,hyperref,amsfonts,pifont}

\usepackage{centernot}

\usepackage[top=1 in, bottom=1 in, left=.9 in, right=.9 in]{geometry}

\newcommand{\captionfonts}{\footnotesize} 
\makeatletter
\long\def\@makecaption#1#2{%
  \vskip\abovecaptionskip
  \sbox\@tempboxa{{\captionfonts #1: #2}}%
  \ifdim \wd\@tempboxa >\hsize
    {\captionfonts #1: #2\par}
  \else
    \hbox to\hsize{\hfil\box\@tempboxa\hfil}%
  \fi
  \vskip\belowcaptionskip}
\makeatother

\def\lsim{ \lower .75ex \hbox{$\sim$} \llap{\raise .27ex
\hbox{$<$}} }
\def\gsim{ \lower .75ex \hbox{$\sim$} \llap{\raise .27ex
\hbox{$>$}} }

\renewcommand{\title}[1]{\vbox{\center\LARGE{#1}}\vspace{5mm}}
\renewcommand{\author}[1]{\vbox{\center#1}\vspace{5mm}}
\newcommand{\address}[1]{\vbox{\center\em#1}}

\newcommand{\starttext}{
\setcounter{footnote}{0}
\renewcommand{\thefootnote}{\arabic{footnote}}}

\newcommand{\be}{\begin{equation}}
\newcommand{\bea}{\begin{eqnarray}}
\newcommand{\eea}{\end{eqnarray}}
\newcommand{\beq}{\begin{equation}}
\newcommand{\ee}{\end{equation}}

\let\oldsqrt\sqrt
\def\sqrt{\mathpalette\DHLhksqrt}
\def\DHLhksqrt#1#2{%
\setbox0=\hbox{$#1\oldsqrt{#2\,}$}\dimen0=\ht0
\advance\dimen0-0.2\ht0
\setbox2=\hbox{\vrule height\ht0 depth -\dimen0}%
{\box0\lower0.4pt\box2}}

\def\sc{\setcounter{equation}{0}}

\begin{document}
\begin{titlepage}

\rightline{}
\bigskip
\bigskip\bigskip\bigskip\bigskip
\bigskip

\centerline{\large \bf {A Quantum Complexity Lowerbound from Differential Geometry}}

\bigskip

\bigskip
\begin{center}

\author{Adam R. Brown}

\address{Google Research (Blueshift), Mountain View, CA 94043, USA}

\address{Stanford Institute for Theoretical Physics and Department of Physics, \\
Stanford University, Stanford, CA 94305, USA}

\end{center}

\begin{center}
\bf     \rm

\bigskip

\end{center}

\begin{abstract}

The Bishop-Gromov bound---a cousin of the focusing lemmas that Hawking and Penrose used to prove their black hole singularity theorems---is a differential geometry result that upperbounds the rate of growth of volume of geodesic balls in terms of the Ricci curvature. In this paper, I apply the Bishop-Gromov bound to Nielsen's complexity geometry to prove lowerbounds on the quantum complexity of a typical unitary. For a broad class of penalty schedules, the typical complexity is shown to be exponentially large in the number of qubits. This technique gives results that are tighter than all known lowerbounds in the literature, as well as establishing lowerbounds for a much broader class of complexity geometry metrics than has hitherto been bounded. For some metrics, I prove these lowerbounds are  tight. This method realizes the original vision of Nielsen, which was to apply the tools of differential geometry to study quantum complexity.

\medskip
\noindent
\end{abstract}

\let\thefootnote\relax\footnotetext{email: \tt{mr.adam.brown@gmail.com}}

\end{titlepage}

\starttext \baselineskip=17.63pt \setcounter{footnote}{0}

\vfill\eject

\tableofcontents

\vfill\eject

\section{Introduction} 

Geometric ideas were  introduced into the study of quantum complexity by Nielsen and collaborators \cite{Nielsen1,Nielsen2,Nielsen3,Nielsen4,NielsenSingleQubit}. Their vision was that by considering a definition of complexity that replaces quantum gates with a smooth path through unitary space, the tools of differential geometry might be brought to bear on proving complexity lowerbounds. So far this vision has not been realized. In this paper, I will use a theorem from differential geometry---the Bishop-Gromov bound \cite{BishopGromov,bishopgromov}---to prove a complexity lowerbound for the complexity geometry. I address the following basic question:
\begin{itemize}
\item[] {\bf Question:} For the complexity geometry on $N$ qubits, does the complexity of a typical operator grow exponentially with $N$? 
\end{itemize}
For one particular complexity geometry---what below we will call the `cliff' metric---this question has already been answered in the affirmative by Nielsen, Dowling, Gu, and Doherty  \cite{Nielsen2}, using a non-geometric technique that leverages prior results from gate complexity. In this paper I will use a tool from differential geometry
to rederive this result in a way that makes no mention of  gates, $\epsilon$-balls, or error budgets,  and improves the exponent. I will argue that my improved lowerbound is tight. Next, I will apply this technique to many different complexity geometries, and show that for a broad class of metrics, many much less complex than the cliff metric, the complexity of a typical unitary is still exponential in $N$.

\subsection{Review of gate complexity and complexity geometry}
The complexity of a transformation quantifies how hard it is to implement. The group of transformations we will be interested in implementing is U($2^N$), the purity-preserving linear functions on $N$ qubits. Let's consider two different sets of primitive operations we might use to build elements of this group, which will yield two different definitions of complexity. \\

\noindent  {\bf Gate Complexity.} For gate complexity, we compile complex unitaries by arranging gates in a circuit. Usually the primitive gates may only act on a small number---often two---qubits at a time. There may either be a discrete set of primitive gates (e.g.~CNOT plus Hadamard plus a random phase are known to be universal \cite{Nielsen:2011:QCQ:1972505}) or a continuous set (for example we may permit any transformation U($2^2$) on the two qubits). The value of the complexity is then the number of primitive gates required to build the target unitary. 

We are interested in the  number of gates required to synthesize typical elements of U($2^N$). For a discrete gate set, we must introduce a tolerance $\epsilon$ if we aspire to reach every element, whereas for a continuous gate set we can  hit every element exactly. Either way, we can derive an exponential lowerbound just by counting. For discrete gates, we count the number of $\epsilon$-balls in U($2^N$), namely $e^{O(\log[\epsilon^{-1}] 4^N)}$, and then count the number of different circuits with $\mathcal{C}$ gates, $e^{O(  \mathcal{C} \log[N])}$; for continuous gate sets we count the number of degrees of freedom in U($2^N$), namely $4^N$, and then count the dimensionality of the space of gates, O(1); in both cases therefore we cannot hope to have fully explored the majority of the unitary group until 
\begin{equation}
\mathcal{C}_\textrm{gates} [ U(2^N) ] \ \gsim \ 4^N \ . \label{eq:gateupperbound}
\end{equation}
For many sets of primitive gates (including the continuous two-qubit gate set U($2^2$)) we can, with a little more work, prove that this complexity lowerbound is approximately tight. For gate complexity the question of the complexity of a typical unitary thus has a simple and settled answer. Not so for complexity geometry. \\

\noindent  {\bf Complexity Geometry.} Complexity geometry introduce a new metric on the unitary group that  stretches directions that are hard to move in, so that complex unitaries are pushed farther away \cite{Nielsen1,Nielsen2,Nielsen3,Nielsen4,NielsenSingleQubit}. The complexity of a unitary may then be defined as the length of the shortest path connecting $U$ to the identity. We thus compile a complex unitary by gliding through U($2^N$) on a continuous path, guided by a (possibly time-dependent) Hamiltonian. 

A complete basis for the Hamiltonians in the tangent space of U$(2^N)$ is given by the generalized-Pauli operators. A $k$-local generalized-Pauli $\sigma_I$ is the product of $k$ single-qubit Pauli operators and $N-k$ identity operators that act on the other qubits. For example $(\sigma_x)_1 \otimes \mathds{1}_2 \otimes  (\sigma_y)_3 \otimes   (\sigma_y)_4  \otimes \mathds{1}_5 \otimes \ldots  \otimes \mathds{1}_N$ is a 3-local generalized Pauli, where the numerical subscript indicates which qubit the operator acts on. There are 
\begin{equation}
\mathcal{N}_k \equiv {N \choose k} 3^k
\end{equation} 
different $k$-local (`weight $k$') generalized Pauli's, for a total of $\sum_{k=0}^N  {N \choose k} 3^k = 4^N$. $\mathcal{N}_k$ is peaked at $k = \frac{3}{4}$. In this basis, the complexity distance between $U$ and $U + d U$ is then defined by \cite{Nielsen1}
\begin{equation}
ds^2 = \sum_{IJ} \textrm{Tr}[i \, dU \, U^\dagger  \sigma_I] \mathcal{I}_{IJ} \textrm{Tr}[i \, dU \, U^\dagger  \sigma_J] \ . \label{eq:definitionofcomplexitygeometry}
\end{equation}
(See also Sec.~1 of \cite{Brown:2019whu} for a pedagogical introduction, and \cite{COMPLEXITYGEOMETRYRECENT} for other recent work.) Here $\mathcal{I}_{IJ}$ is a positive-definite matrix. Were $\mathcal{I}_{IJ}$ the identity matrix, $\mathcal{I}_{IJ} = \delta_{IJ}$, this would recover the standard `bi-invariant' inner-product metric on U$(2^N)$; in the inner-product metric the distance $s$ between two unitaries is given by $\cos s = |\textrm{Tr}[U_1 U_2^{\dagger}] |$, where we have normalized the trace so that the maximum separation (the `diameter') is $\pi$. In the inner-product metric, it is equally `easy' to move in any direction. In the complexity geometry, we change $\mathcal{I}_{IJ}$ so that while it may still be inexpensive to move in some directions (e.g.~in the directions of Hamiltonians that act nontrivially on only two qubits), other directions may be assigned a large `penalty factor' $\mathcal{I}_{IJ}$ to reflect that moving in highly non-local directions is hard. The $\mathcal{I}_{IJ}$s we will consider in this paper will be diagonal in the generalized-Pauli basis, and will have a diagonal value that depends only on the $k$-locality, so that the metric is fully specified by the `penalty schedule'  $\mathcal{I}_k$ that assigns a penalty factor to each value of $k \leq N$. The infinitesimal distance in off-diagonal directions is then determined by Pythagoras' theorem, since Eq.~\ref{eq:definitionofcomplexitygeometry} has an L2 norm. Finally, notice that the metric  Eq.~\ref{eq:definitionofcomplexitygeometry} is `right-invariant', which means the distance from $\mathds{1}$ to $U$ is the same as the distance from ${U}_R$ to ${U} U_R$, as a consequence of which the complexity geometry is homogeneous. 

We want to know the complexity of a typical operator, as a function of $N$ and the penalty schedule $\mathcal{I}_k$. Unlike for gate complexity, we cannot just count. We can't count points because the number of points reachable by even an arbitrarily short path is uncountable. And we can't even count dimensions, since there is no restriction on how often the path may change direction, so the dimensionality of the space of even arbitrarily short paths is infinity. 
To lowerbound the complexity of a typical unitary, we cannot use counting. Instead we must measure volume. But first let's note some useful lemmas.

The first useful lemma is that the  gate complexity \emph{upper}bounds the Nielsen complexity. This is because the complexity geometry definition is  \emph{more} permissive than the gate definition. For example, any $k$-qubit gate can be implemented by evolving with an at-most-$k$-local Hamiltonian for an inner-product distance at most $\pi$. So even if we make a more permissive definition of gate complexity that allows us to deploy any U($2^k$) $k$-qubit gate at a cost of $\sqrt{ \mathcal{I}_k}$ per gate, the gate complexity upperbounds the geometric complexity 
\begin{equation}
\mathcal{C}_{\textrm{geom.,} \mathcal{I}_k }   [ U ] \ \leq \ \pi \, \mathcal{C}_{\textrm{gates}, \mathcal{I}_k}  [ U ]  \ . \label{eq:geomlessthangatesobvs}
\end{equation}

The second useful lemma relates distances in two different complexity geometries. The more expensive penalty schedule must have concomitantly longer distances, so 
\begin{equation}
\forall k, \mathcal{I}_k \geq \bar{\mathcal{I}}_k \ \ \rightarrow \ \  \mathcal{C}_{\textrm{geom,} \mathcal{I}_k }   [ U ] \ \geq \ \mathcal{C}_{\textrm{geom,} \bar{\mathcal{I}}_k }[U]  \ . \ \ \  \label{eq:frankensteinlicense}
\end{equation}
As a special case, increasing every $\mathcal{I}_k$ in a penalty schedule to be equal to  the maximum penalty gives a rescaled version of the inner-product metric, and so upperbounds
\begin{equation}
\mathcal{C}_{\textrm{geom,} \mathcal{I}_k }   [ U ] \ \leq \ \pi \, \textrm{max}_{k} \sqrt{ {\mathcal{I}}_k}. \label{eq:geomlessthanmaxIobvs}
\end{equation}

\subsection{The Bishop-Gromov bound} 
The volume of a geodesic ball of radius $\mathcal{C}$ in $d$-dimensional hyperbolic space is given by  $\textrm{volume}_{\textrm{BG}}(\mathcal{C})$ where, for unit vectors $\vec{X}$, 
\begin{equation}
 \textrm{volume}_{\textrm{BG}}(\mathcal{C}) \equiv  \Omega_{d-1} \, \int_0^\mathcal{C} d\tau \left( \sqrt{ \frac{d-1}{-    \textrm{min}_{\vec{X}}  [\mathcal{R}_{ \mu \nu } X^\mu X^\nu]} } \sinh \left[   \frac{\sqrt{ -  \textrm{min}_{\vec{X}}[ \mathcal{R}_{ \mu \nu } X^\mu X^\nu]}}{\sqrt{d-1}}  \tau \right] \right)^{d-1} . 
\end{equation}
Since hyperbolic space is isotropic, the `min' here is redundant as the Ricci curvature is the same in all directions (and so  $\mathcal{R}_{ \mu \nu } X^\mu X^\nu$ is independent of $\vec{X}$). But the Bishop-Gromov theorem \cite{BishopGromov,bishopgromov} says that this same expression upperbounds the volume of geodesic balls in any $d$-dimensional homogeneous space, even when it's not isotropic 
\begin{equation}
\textrm{BG theorem:  } \ \ \ \ \   \textrm{volume}_{\textrm{ball}}(\mathcal{C}) \ \leq \ \textrm{volume}_{\textrm{BG}}(\mathcal{C}) \ . \ \ \ \ \ \  \label{eq:BGBGBG}
\end{equation}
Though calculating the exact volume of the geodesic ball is hard, the BG theorem gives a simple upperbound
 in terms of a single local geometric quantity. 

(For mathematicians, the Bishop-Gromov bound is a venerable `comparison theorem' of differential geometry that needs no introduction, but physicists may find it helpful to think of it as arising from the $d+0$-dimensional Raychaudhuri equation \cite{Raychaudhuri:1953yv} 
\begin{equation}
{\theta}' + \frac{1}{d-1} \theta^2 - \omega^2 =  -  \sigma^2  - \mathcal{R}_{\mu \nu}X^{\mu} X^{\nu} \ , \label{eq:RayRayRaychaudhuri}
\end{equation}
where if you don't know what those letters mean\footnote{Or if they're all Greek to you.}  
please don't read this paragraph. The $3+1$ version of this focussing equation was most famously deployed by Penrose and Hawking \cite{Penrose:1964wq}, who assured a non-negative $\mathcal{R}_{\mu \nu}X^{\mu}X^{\nu}$ by applying Einstein's equations and then imposing the Weak Energy Condition, and then used the negativity of the right-hand side to 
upperbound the expansion of lightsheets and show that `trapped surfaces' inevitably hit black hole singularities. To derive the BG bound, one lowerbounds $\mathcal{R}_{\mu \nu}X^{\mu} X^{\nu}$ by assumption, puts $\omega = 0$ since a geodesic ball has zero angular momentum, and then ignores the shear $\sigma$ since it can only make the  ball grow slower. Integrating along geodesics leaving the origin then yields Eq.~\ref{eq:BGBGBG}.) 

\subsection{Proof strategy overview}
Our proof strategy is, schematically, 
\begin{equation}
\textrm{bound curvature below} \ \rightarrow \  \textrm{bound volume$_\textrm{ball}(\mathcal{C})$ above}\  \rightarrow \ \textrm{bound  complexity below},
\end{equation}
as we will now explain. The complexity metric has an infinite number of points but only a finite (though generally double-exponentially large) volume, 
\begin{equation}
\textrm{volume}_{\mathcal{I}_k} (U(2^N))  = \int \sqrt{ |g|} = \omega_d \sqrt{ \prod_{k \, \leq \, N}  (\mathcal{I}_k)^{ { N \choose k} 3^k } } \ .  \label{eq:volumeofIk}
\end{equation}
Here $\omega_d$ is the volume of the unit $d=4^N$-dimensional unitary group U($2^N$) with the inner-product metric,  $\mathcal{I}_k = 1$.
 Since the complexity of an operator is defined as its geometric distance from the identity, the volume of unitaries with complexity less than $\mathcal{C}$ is simply the volume of the geodesic ball with radius $\mathcal{C}$.  We cannot hope to be able to synthesize the median unitary until this geodesic ball has engulfed (half of) the space 
  \begin{equation}
 \textrm{volume}_\textrm{ball}(\mathcal{C}_\textrm{typical}) = \frac{1}{2} \textrm{volume}_{\mathcal{I}_k} (U(2^N)) \ .  \label{eq:typicalunitarydefinition}
 \end{equation}
Combining Eqs.~\ref{eq:BGBGBG}, \ref{eq:volumeofIk}, and \ref{eq:typicalunitarydefinition}, we will lowerbound $\mathcal{C}_\textrm{typical}$ using
  \begin{equation}
 \textrm{volume}_{\textrm{BG}}(\mathcal{C}_\textrm{typical})  \ \geq \  \textrm{volume}_\textrm{ball}(\mathcal{C}_\textrm{typical}) = \frac{1}{2} \omega_d \sqrt{ \prod_{k \, \leq \, N}  (\mathcal{I}_k)^{ { N \choose k} 3^k } }  \ . \label{eq:mainequation}
 \end{equation}

\subsection{The Ricci curvature}  \label{sec:riccicurvatureapproximate}
To evaluate the Bishop-Gromov bound, we must calculate the most negative component of the Ricci curvature. The curvature of right-invariant metrics was investigated by Milnor \cite{Milnor} (note mathematicians use the mirror convention and call them `left-invariant'). As recounted in Appendix A, the Ricci curvature is diagonal in the generalized Pauli basis and given by 
\begin{equation}
\mathcal{R}_{\sigma_I}^{\ \sigma_I} =  \sum_{\sigma_J} \frac{(\textrm{Tr}([\sigma_I,\sigma_J]^2))^2}{4}   \frac{ \mathcal{I}_{\sigma_I}^2 - (\mathcal{I}_{[\sigma_I,\sigma_J]} - \mathcal{I}_{\sigma_J})^2 }{4\mathcal{I}_{\sigma_I} \mathcal{I}_{\sigma_J} \mathcal{I}_{[\sigma_I,\sigma_J]}} 
 \, . \label{eq:tobereferredtolater}
\end{equation}
Define $\textrm{\#overlap}_\textrm{same}$ as the number of qubits to which $\sigma_I$ and $\sigma_J$ assign the same SU(2) Pauli ($\sigma_x$, $\sigma_y$ or $\sigma_z$) and $\textrm{\#overlap}_\textrm{diff.}$ as the number of qubits to which they both assign a Pauli (i.e.~not $\mathds{1}$) but the Pauli's are different. Then  $\frac{1}{4}\textrm{Tr}([\sigma_I,\sigma_J]^2)$ will be 1 if $\textrm{\#overlap}_\textrm{diff.}$ is odd and 0 otherwise.  When  the commutator is nonzero it is another generalized Pauli with weight 
\begin{equation}
\textrm{weight}( [\sigma_I, \sigma_J])  =   \textrm{weight}(\sigma_I) + \textrm{weight}( \sigma_J ) -  \textrm{\#overlap}_\textrm{diff.} -  2 \textrm{\#overlap}_\textrm{same} \ . \label{eq:weightweightweight}
\end{equation}
These formulae, plus some combinatorics, are sufficient to calculate the Ricci curvature for every direction and every penalty schedule. 

\subsection{Approximations}
The Bishop-Gromov technique can give a precise lowerbound, but in this paper we will extract only the exponential behavior in $N$. This will allow us to make a number of simplifications. 

First, since we are interested in a lowerbound, we will drop the positive term in Eq.~\ref{eq:tobereferredtolater}. Since $\frac{1}{2}[\sigma_I , (\frac{1}{2}[\sigma_I, \sigma_J])] = \sigma_J$ we can reorder the sum to write 
\begin{equation}
\mathcal{R}_{\sigma_I}^{\ \sigma_I} \ \geq \  -  2 \hspace{-8mm}   \sum_{\substack{\sigma_J \textrm{ with } \\  w(\sigma_J) < w([\sigma_I,\sigma_J])}}  \hspace{-8mm} \frac{(\textrm{Tr}([\sigma_I,\sigma_J]^2))^2}{4}     \frac{ (\mathcal{I}_{[\sigma_I,\sigma_J]} - \mathcal{I}_{\sigma_J})^2 }{4\mathcal{I}_{\sigma_I} \mathcal{I}_{\sigma_J} \mathcal{I}_{[\sigma_I,\sigma_J]}}   \, . \label{eq:tobereferredtolater2}
\end{equation}

Second, to make the double-exponentially large terms more manageable, we can take the $4^N$th root of Eq.~\ref{eq:mainequation}, which to good approximation becomes 
\begin{equation}
\left( \Omega_{4^N} \right)^\frac{1}{4^N} L  \sinh \left[   \frac{  \mathcal{C_\textrm{typical}}}{L}  \right]  \  \geq \  \Bigl( \frac{1}{2} \Bigl)^\frac{1}{4^N}   \left( \omega_{4^N} \right)^\frac{1}{4^N} \sqrt{\mathcal{I}_\textrm{av.} } \label{eq:BishopGromovsimplified}
\end{equation} 
where $L^2 \equiv  \frac{4^N}{-    \textrm{min}_{\vec{X}}  [\mathcal{R}_{ \mu \nu } X^\mu X^\nu]} $ and $\mathcal{I}_\textrm{av.}$ is the geometric mean of the penalty factors. The $( \frac{1}{2})^\frac{1}{4^N}$ term is very close to one and can be dropped:  in this high-dimensional negatively curved space,  geodesic balls grow so rapidly that the bounds we can place on the \emph{typical} complexity are very close to the bounds we can place on the \emph{worst-case} complexity (the `diameter'), and henceforth we'll elide the two. We can also neglect the difference between the volume of the unit unitary $\omega_{4^N}$ and the volume of the unit sphere $\Omega_{4^N}$, since the ratio is the trifling $\bigl( \frac{\omega_{4^N}}{\Omega_{4^N}}\bigl)^\frac{1}{4^N} \sim  \frac{3}{4}$.  Eq.~\ref{eq:BishopGromovsimplified} then implies 
\begin{equation}
\boxed{ \mathcal{C}_\textrm{typical}   \  \  \gsim  \  \ \textrm{min}\bigl[\sqrt{\mathcal{I}_\textrm{av.} } , L \bigl]   \ \equiv \ \textrm{min}\Bigl[\sqrt{\mathcal{I}_\textrm{av.} } , \sqrt{ \frac{4^N}{-    \textrm{min}_{\vec{X}}  [\mathcal{R}_{ \mu \nu } X^\mu X^\nu]} } \Bigl] }   \ . \label{eq:BishopGromovsimplified2}
\end{equation}

. 
\section{Complexity Lowerbounds} 
\sc
\subsection{The cliff metric}
The first and simplest application of the Bishop-Gromov technique will be to a case already considered by Nielsen, Dowling, Gu, and Doherty  \cite{Nielsen2}, 
\begin{equation}
\textrm{cliff schedule: } \ \ \ \mathcal{I}_1 = \mathcal{I}_2 = 1 \ \ \ \mathcal{I}_{k \geq 3} = \mathcal{I}_\textrm{cliff} = q \ . \label{eq:cliffschedule}
\end{equation}
To lowerbound the diameter, we must lowerbound the Ricci curvature.  The terms in Eq.~\ref{eq:tobereferredtolater2} can only be nonzero when weight$(\sigma_J) = 2$ and weight$([\sigma_I,\sigma_J]) > 2$ .  Since there are only $\mathcal{N}_2 = {N \choose 2}3^2$ two-local generalized Pauli's in total, the curvature is certainly lowerbounded by 
\begin{equation}
\mathcal{R}_{\sigma_I}^{\ \sigma_I} \ \geq \  - 2  {N \choose 2} 3^2 \frac{ (q - 1)^2 }{4 q \mathcal{I}_{\sigma_I} }  \ . 
\end{equation} 
This equation tells us the most negative components of the Ricci curvature are in the 2-local directions. This is easy to understand: the sectional curvature is negative when two easy directions commute to a hard direction, and indeed we see in this calculation that the only negative contributions to the Ricci tensor are when a two-local direction (easy) commutes with another two-local direction (easy) to make a three-local direction (hard). This gives\footnote{Note that this approximate expression is consistent with the exact answer calculated by Dowling and Nielsen
in Appendix A of \cite{Nielsen4}, namely  $\mathcal{R}_{2}^{\ 2} = 
 -24(N-2)q +8(6N-11) 
 + \left(\frac{1}{2} 4^N -8(3N-5)
    \right)q^{-2} $.}
\begin{equation}
\textrm{min}_{\vec{X}}  [\mathcal{R}_{ \mu \nu } X^\mu X^\nu]  = \mathcal{R}_{2}^{\ 2} \ \geq \  -  2  {N \choose 2} 3^2 \frac{ (q - 1)^2 }{4 q  }  \  \sim \   - q \ ,
\end{equation}
where in the final term we have dropped all terms sub-exponential in $N$. On the other hand, the geometric mean of the penalty factors $\mathcal{I}_{\textrm{av.}}$ is exponentially close to $q$. The Bishop-Gromov bound Eq.~\ref{eq:BishopGromovsimplified2} thus yields 
\begin{equation}
 \mathcal{C}_\textrm{typical}   \  \  \gsim  \ \ \textrm{min}\Bigl[ \sqrt{q} , \sqrt{ \frac{4^N}{q} } \Bigl] \ . \label{eq:attempt1} 
\end{equation} 
(We use `$\gsim$' not `$\geq$' since we have neglected terms that are not exponential in $N$.) 
This is equal to $\sqrt{q}$ up to $q = {2^N}$, and then decreases again. Even though the lowerbound gets lower for $q > {2^N}$, the complexity itself does not. Eq.~\ref{eq:frankensteinlicense} tells us that the complexity for $q > {2^N}$ is lowerbounded by the complexity for $q = {2^N}$, so Eq.~\ref{eq:attempt1}  becomes 
\begin{equation}
 \mathcal{C}_\textrm{typical}   \  \  \gsim  \ \ \textrm{min}\Bigl[ \sqrt{q} , {2^\frac{N}{2} } \Bigl] \ . \label{eq:attempt1a} 
\end{equation} 
For exponentially large $q$ this gives an exponentially large lowerbound. \\

\noindent The step from Eq.~\ref{eq:attempt1} to \ref{eq:attempt1a} was the first example of a trick we will use repeatedly in this paper: even if a metric is too highly curved for the BG bound to have purchase, we may be able to lowerbound the complexity by finding an easier (lower $\mathcal{I}_k$) metric that is gently curved. Indeed, we can  immediately use this trick to improve Eq.~\ref{eq:attempt1a}. The negative curvature was driven by two 2-local directions commuting to a 3-local direction, giving a sectional curvature  
$- \frac{\mathcal{I}_3}{\mathcal{I}_2^{\, 2}} = -q$. If we made an easier metric with an intermediate step, $\mathcal{I}_3 = \sqrt{q}$, then the 2-2 sections  (and 2-3 sections) would contribute only $- \frac{\mathcal{I}_3}{\mathcal{I}_2^2} = - \frac{\mathcal{I}_4}{\mathcal{I}_2 \mathcal{I}_3} = -\sqrt{q}$. Adding more such steps and optimizing the staircase leads, as we will see in Eq.~\ref{eq:binomialresult}, to 
\begin{equation}
 \mathcal{C}_\textrm{typical}   \  \  \gsim   \ \ \textrm{min}\Bigl[ \sqrt{q} , {4^\frac{N}{2} } \Bigl] \ . \label{eq:attempt2} 
\end{equation}  
We can combine this with the upperbounds given by Eqs.~\ref{eq:geomlessthangatesobvs} and \ref{eq:geomlessthanmaxIobvs} to yield
\begin{equation}
\textrm{min}\Bigl[ \sqrt{q} , {2^N } \Bigl] \ \leq \  \mathcal{C}_\textrm{typical}  \ \leq \ \textrm{min}[\sqrt{q},4^N] \ . \  \label{eq:upperlowerboundscliffmetricdiameterqubits1}
\end{equation}
For $q < 4^N$ we have fixed (the exponential part) of the complexity exactly: it's $\sqrt{q}$. 
     \begin{figure}[htbp] 
    \centering
    \includegraphics[width=5.5in]{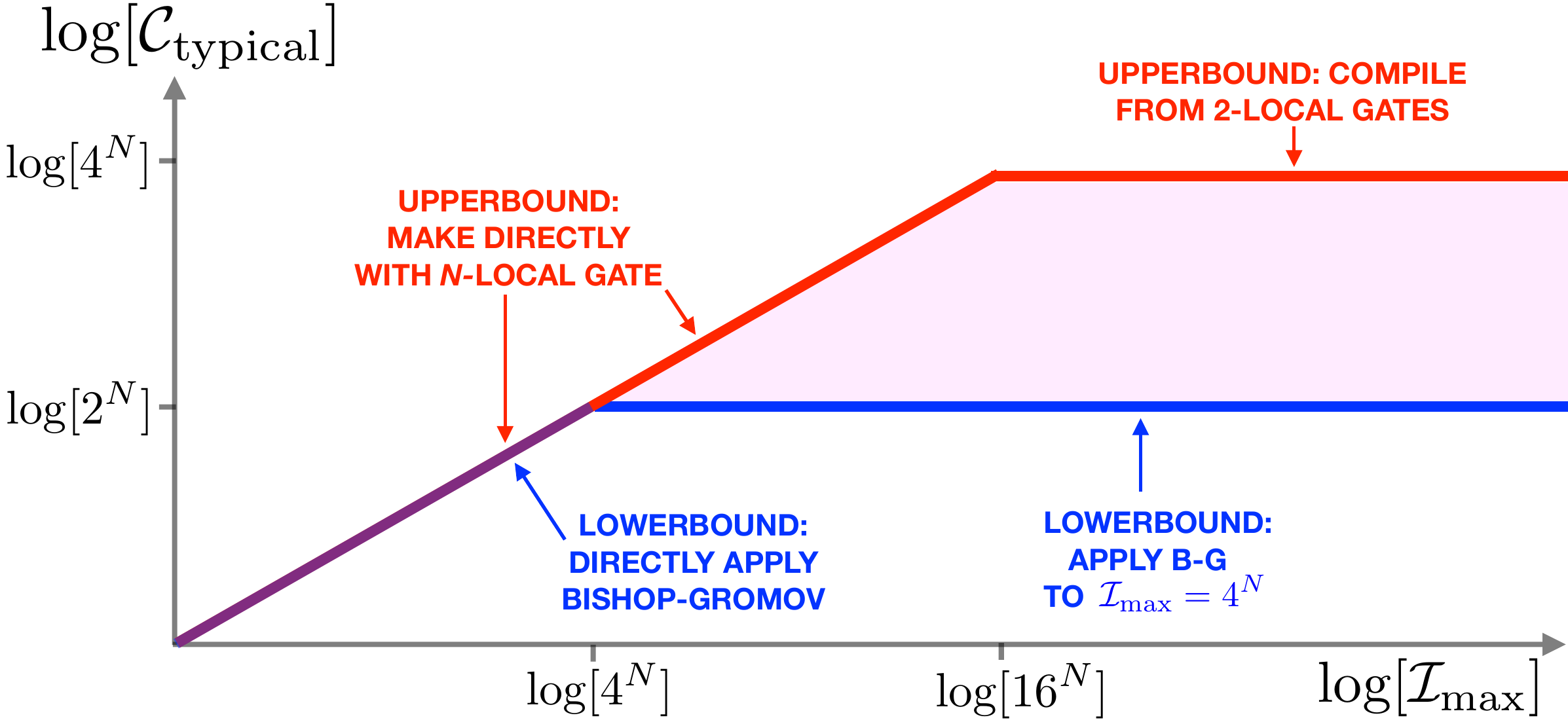} 
    \caption{The (exponential part of the) diameter of the cliff metric for $N$ qubits with  $\mathcal{I}_1 = \mathcal{I}_2 =  1$, $\mathcal{I}_{k \, \geq \, 3} = \mathcal{I}_\textrm{max}$, given by Eq.~\ref{eq:upperlowerboundscliffmetricdiameterqubits1}. For $\mathcal{I}_\textrm{max} < 4^N$ the (exponential part of the) diameter is exactly $\sqrt{\mathcal{I}_\textrm{max}}$, for $\mathcal{I}_\textrm{max} > 4^N$ we can bound the diameter above and below: the diameter is somewhere in the shaded region. In Sec.~\ref{sec:discussion} we will conjecture that the correct answer is given by the blue line and the Bishop-Gromov lowerbound is tight. }
        \label{fig:boundsdiametercliffqubit}
 \end{figure}

\subsection{The delayed-cliff metric} 
The delayed-cliff schedule is cheap for all $k$s up to some $ K < \frac{3}{4} N$, 
\begin{equation}
\textrm{delayed-cliff schedule: } \ \ \ \mathcal{I}_{k \leq K} = 1 \ \ \ \mathcal{I}_{k > K} = \mathcal{I}_\textrm{cliff} = q \ . \label{eq:delaycliffschedule}
\end{equation}
The only non-zero contributions to the sum Eq.~\ref{eq:tobereferredtolater2} will be when 
$w(\sigma_J) \leq K < w([\sigma_I,\sigma_J]) $. This means there cannot be more terms in the sum than there are $\sigma_J$s with weight less than $K$, so a crude lowerbound on the Ricci curvature is that  
\begin{equation}
\textrm{min}_{\vec{X}}  [\mathcal{R}_{ \mu \nu } X^\mu X^\nu] \  \   \ >  \ -\left(  \sum_{k=0}^K \mathcal{N}_k \right) \frac{(q-1)^2}{q} \ \sim \ - \mathcal{N}_K \, q  \ . \label{eq:minRdelayedcliff} 
\end{equation}
So long as $\frac{K}{N} < \frac{3}{4}$, then to excellent approximation $\mathcal{I}_{\textrm{av.}} = q$, and an (unoptimized) complexity lowerbound is given by combining Eqs.~\ref{eq:BishopGromovsimplified2} and \ref{eq:minRdelayedcliff}, 
\begin{equation}
 \mathcal{C}_\textrm{typical}   \  \  \gsim  \ \ \textrm{min}\Bigl[ \sqrt{q} , \sqrt{ \frac{4^N }{ \mathcal{N}_K \, q} }  \Bigl] \ \ \rightarrow  \ \ \mathcal{C}_\textrm{typical}   \  \  \gsim  \ \ \textrm{min}\Bigl[ \sqrt{q} ,  \frac{2^\frac{N}{2}  }{ (\mathcal{N}_K)^\frac{1}{4} }   \Bigl] \ . \label{eq:randomizedcliff}
\end{equation}
This equation would still hold for a `symmetric  delayed-cliff schedule' that also makes cheap the $\mathcal{N}_K$ directions with the \emph{largest} values of $k$. This implies
\begin{eqnarray}
& & \textrm{\emph{If all directions are cheap $\mathcal{I}_k = 1$ except those with $\frac{3}{4} - \epsilon < \frac{k}{N} < \frac{3}{4} + \epsilon$ are exponentially}} \nonumber \\
& &\textrm{\emph{expensive, then for all $\epsilon> 0$ the typical unitary will be exponentially complex.}}
\end{eqnarray}
The point is that the number of generalized Pauli's, $\mathcal{N}_k \equiv {N \choose k}3^k$, is so sharply peaked at $k = \frac{3}{4}N$ (thanks to the central limit theorem) that even though the range $0.749N<k<0.751N$ contains only a small fraction of the values of $k$, it contains almost all the generalized Pauli's.

\subsection{Pick `$\mathcal{N}$' Mix}
Consider a `randomized' version of the delayed cliff metric, in which we pick \emph{any} $\mathcal{N}$ of the generalized Pauli's to be easy $\mathcal{I}= 1$, and all others to be hard. Nothing about the argument that lead to Eq.~\ref{eq:randomizedcliff} relied on the easy directions sharing a particular $k$-locality, and so the same bound still applies\footnote{Though I expect the \emph{optimized} bound for the delayed cliff is slightly tighter than the \emph{optimized} bound for a random assortment with the same number of cheap directions,  because if you respect $k$-locality you get fewer distinct commutators than you do for random assortments, since $\mathcal{N}_{2k} < \mathcal{N}_{k}^{ \ 2}$. This is what $k$-locality `buys' you.}. So long as  $4^N/\mathcal{N}$ is still exponentially large, and so long as the hard directions are exponentially expensive, the typical complexity will still be exponential in $N$.

\subsection{The exponential metric} \label{sec:exponentialmetric}
The exponential schedule (discussed in \cite{Brown:2017jil,Brown:2021rmz}) is defined by 
\begin{equation}
\textrm{exponential schedule: } \ \ \ \mathcal{I}_{k} = x^{2k}  \ . \label{eq:exponentialschedule} 
\end{equation}   
We know immediately that for all $x>1$ the complexity of a typical unitary will be exponential in $N$. This follows from the fact that the exponential schedule clearly upperbounds a delayed cliff schedule with exponentially large $q$ (e.g. $K = \frac{1}{2} N \ \&  \ q = x^N$).

Let's strive for a tighter bound.  Unfortunately the exponential metric is too highly curved for a direct application of the Bishop-Gromov bound to be useful. Instead, we will first lowerbound the complexity of the (strictly easier) `truncated exponential schedule', 
\begin{equation}
\textrm{truncated exponential schedule: } \ \ \ \mathcal{I}_{k} = \textrm{min}\Bigl[ x^{2k} , x^{2 \bar{k}} \Bigl] \ , 
\end{equation}
and then use Eq.~\ref{eq:frankensteinlicense}. 
The most negative contribution to the Ricci curvature comes when $w(\sigma_I) =2$, $w(\sigma_J) = \bar{k} - 1$, and $w([\sigma_I,\sigma_J]) = \bar{k} $ giving 
\begin{equation}
\textrm{min}_{\vec{X}}  [\mathcal{R}_{ \mu \nu } X^\mu X^\nu]  \ = \ \mathcal{R}_{2}^{\ 2} \  \ \gsim \  \ - \mathcal{N}_{\bar{k}-1}   \frac{\mathcal{I}_{\bar{k}}   }{\mathcal{I}_{\bar{k}-1} \mathcal{I}_{2}  } \   \ \sim \ - \mathcal{N}_{\bar{k}} \ . 
\end{equation}
Since $\mathcal{N}_k$ is sharply peaked around $k = \frac{3}{4} N$, to excellent approximation $\mathcal{I}_{\textrm{av.}} = x^{2\bar{k}}$, and Eq.~\ref{eq:BishopGromovsimplified2} gives 
\begin{equation}
 \mathcal{C}_\textrm{typical}   \  \  \gsim  \  \ \textrm{min} \Bigl[     x^{\bar{k}} , \sqrt{ \frac{4^N}{    \mathcal{N}_{\bar{k}} } }  \Bigl] \ . 
\end{equation}
Optimizing this expression over $\bar{k}$ gives $\mathcal{N}_{\bar{k}} \,  x^{2 \bar{k}}  = 4^N$ and so 
\begin{equation}
 \mathcal{C}_\textrm{typical}   \  \  \gsim  \  \    x^{\bar{k}}  \ \ \textrm{where }  \mathcal{N}_{\bar{k}} \,  x^{2 \bar{k}}  = 4^N .  \label{eq:exponentialfinalanswer}
\end{equation}
For example, for $x=4$ this evaluates to  $\mathcal{C}_\textrm{typical} \geq 4^{(0.349...) N}$.

\subsection{The binomial schedule } \label{sec:binomialschedule}
Next let's consider a penalty schedule for which the $k$-local penalty factor is given by (a power of) the number of $k$-local directions, 
\begin{equation}
\textrm{binomial schedule: } \ \ \ \mathcal{I}_{k} = (\mathcal{N}_k)^{ \alpha} \equiv  \left( { N \choose k}3^k \right)^\alpha \ . \label{eq:binomialschedule} 
\end{equation}   
We know immediately that for all $\alpha> 0$ the complexity of a typical unitary will be exponential in $N$. This follows from the fact that the binomial schedule clearly upperbounds a `symmetric delayed cliff' schedule with exponentially large $q$. 
 To get a tighter lowerbound, we should consider the `truncated binomial schedule'
  \begin{equation}
\textrm{truncated binomial schedule: } \ \ \ \mathcal{I}_{k} = \textrm{min}\Bigl[( \mathcal{N}_k)^{\alpha} , (\mathcal{N}_{\bar{k}})^{\alpha} \Bigl] \ ,
\end{equation}
for some $\bar{k}$. Using Eqs.~\ref{eq:weightweightweight} \& \ref{eq:tobereferredtolater2}, it is straightforward combinatorics to show that the most negative components of the Ricci tensor are in two-local directions, and that the $\sigma_J$ that make the largest contribution are those for which $w(\sigma_J) = \bar{k}-1$ and  $w([\sigma_I,\sigma_J]) = \bar{k}$,  giving 
\begin{equation}
\textrm{min}_{\vec{X}}  [\mathcal{R}_{ \mu \nu } X^\mu X^\nu]  = \mathcal{R}_{2}^{\ 2} \ \gsim \ - \mathcal{N}_{\bar{k}-1}   \frac{\mathcal{I}_{\bar{k}}  }{\mathcal{I}_{\bar{k}-1} \mathcal{I}_{2}  }   \ \sim - \mathcal{N}_{\bar{k}} \ . 
\end{equation}
Since $\mathcal{N}_k$ is sharply peaked around $k = \frac{3}{4} N$, to excellent approximation $\mathcal{I}_{\textrm{av.}} = (\mathcal{N}_{\bar{k}})^\alpha$, and Eq.~\ref{eq:BishopGromovsimplified2} gives 
\begin{equation}
 \mathcal{C}_\textrm{typical}   \  \  \gsim  \  \ \textrm{min} \Bigl[ \sqrt{      \mathcal{N}_{\bar{k}}^{{\alpha} } } , \sqrt{ \frac{4^N}{    \mathcal{N}_{\bar{k}} } }  \Bigl] \ . 
\end{equation}
Optimizing this expression over $\bar{k}$ gives $\mathcal{N}_{\bar{k}}^{1 + \alpha}  = 4^N$ and so 
\begin{equation}
 \mathcal{C}_\textrm{typical}   \  \  \gsim  \  \    2^{\frac{\alpha N }{1 + \alpha} } \ .  \label{eq:binomialresult}
\end{equation}
Taking $\alpha$ large\footnote{Technically we must also multiply the $\mathcal{I}_k$ by $\mathcal{N}_2^{- \alpha}$ to keep $\mathcal{I}_2 = 1$, though this factor is too puny to matter.} recovers the cliff schedule Eq.~\ref{eq:cliffschedule}, except with sufficient tapering at small $k$ to improve the limit to that quoted in Eq.~\ref{eq:attempt2}. (We could have reached the same conclusion by taking $x$ large in Eq.~\ref{eq:exponentialfinalanswer}.)

\section{Discussion} \label{sec:discussion} \sc
In this paper I used the Bishop-Gromov theorem to lowerbound the complexity of a typical unitary in the complexity geometry, for a range of penalty schedules. We saw that the Bishop-Gromov technique was able to prove a lowerbound of $4^{N/2}$ for the cliff schedule, as well as lowerbounds for the other schedules that were also exponential in $N$. For the special case of the cliff metric (and unlike for the other metrics), there was already an exponential lowerbound known due to the work of Nielsen, Dowling, Gu, and Doherty  \cite{Nielsen2}, namely $4^{N/3}$. Both methods are non-constructive, in the sense that even though both assure you that almost all unitaries are highly complex, neither presents you with a certified hard unitary. However my method is non-constructive \emph{even given an oracle for gate complexity}, whereas Ref.~\cite{Nielsen2} proved that if a unitary is exponentially hard to approximate in gate complexity, it is also exponentially hard to reach in the cliff metric. 

One of the other metrics for which we were able to prove an exponential complexity lowerbound was the `exponential' metric, discussed in Sec.~\ref{sec:exponentialmetric}. This confirms part of a conjecture made in \cite{Brown:2017jil}. Indeed, according to the strong form of the conjectures made in \cite{Brown:2021rmz}, all schedules harder than the `critical schedule' should have the same diameter. This paper has provided a consistency check on these ideas by showing that there is a large class of schedules all of which give exponential diameters.

As well as bounding how complex typical unitaries are, the Bishop-Gromov theorem may also be used to quantify how atypical are low-complexity unitaries. 
The Bishop-Gromov theorem `natively' upperbounds the volume of unitaries with complexity less than a given value, but it can also be used to give a lowerbound, because of the theorem that \cite{BishopGromov,bishopgromov}
\begin{equation}
\frac{d}{d \mathcal{C}} \frac{ \textrm{volume}_\textrm{ball.}(\mathcal{C})}{ \textrm{volume}_\textrm{BG}(\mathcal{C})}  \leq 0 \ . 
\end{equation}
This inequality means that if the growth of the geodesic ball ever falls behind the Bishop-Gromov pace, it can never catch up. Since we know that the geodesic ball must have engulfed the whole space by the time the diameter is reached, and since we can upperbound the diameter (see e.g.~Eq.~\ref{eq:upperlowerboundscliffmetricdiameterqubits1}), this allows us to lowerbound the volume growth. \\

Let's make a mathematical conjecture. In Fig.~\ref{fig:boundsdiametercliffqubit}, we saw that for the cliff metric with $q<4^N$ the BG bound was able to determine (the exponential part of) the diameter exactly, whereas for $q>4^N$ there was a gap between the upperbound and the BG lowerbound. My conjecture is that the Bishop-Gromov lowerbound is tight: 
\begin{equation}
\textrm{ {\bf conjecture}: \ \ the infinite-cliff metric ($\mathcal{I}_{k \leq 2} =1$, $\mathcal{I}_{k \geq 3} = \infty$) has diameter $2^N$} \ . \label{eq:conjecture}
\end{equation}
Let's describe two pieces of circumstantial evidence in favor of this conjecture.

The first piece of evidence is that the BG bound cannot be tightened further. In Sec.~\ref{sec:binomialschedule}, we showed that by using a `staircase' technique, we could enhance the BG lowerbound on the complexity of the cliff metric from the naive $2^{\frac{N}{2}}$ to the improved $2^N$. This is as good as it gets: there isn't an even better shape for the staircase that gives an even higher lowerbound. Eq.~\ref{eq:BishopGromovsimplified2} tells us that to get a lowerbound above $2^N$ we'd need both $\textrm{min}_{\vec{X}}  [\mathcal{R}_{ \mu \nu } X^\mu X^\nu]$ to be exponentially small (or positive) and $\mathcal{I}_{\textrm{max}} > 4^N$. However, if $\mathcal{I}_2 = 1$ and $\mathcal{I}_{\textrm{max}} > 4^N$, then there must exist $k$ such that $\mathcal{I}_k> 4 \mathcal{I}_{k-1} $. Eq.~\ref{eq:tobereferredtolater2} tells us that this value of $k$ will make at least an O(1) negative contribution to $\mathcal{R}_{2}^{\ 2}$. To turn this into a rigorous proof we'd need to confirm that the positive terms in Eq.~\ref{eq:tobereferredtolater} cannot fully cancel these negative contributions, and thereby give an upperbound on the most negative component of the Ricci curvature when there is a sufficiently large ratio between $\mathcal{I}_2$ and $\mathcal{I}_\textrm{max}$.

For the second piece of circumstantial evidence, consider a new definition of complexity I'll call zig-zag$_\delta$ complexity. In zig-zag$_\delta$ complexity you move in piecewise-linear segments. Within each segment, you move with a fixed Tr$H^2 \leq 1$ time-independent  2-local Hamiltonian, for an inner-product distance $\delta$. Then for the next segment you pick some new fixed 2-local Hamiltonian. The complexity of a path is defined as $\mathcal{C} = S \delta$, where $S$ is the number of segments. This definition interpolates between the infinite-cliff complexity geometry when $\delta = 0$, and a model that is only somewhat more powerful than the gate definition when $\delta = \pi$.  To reach every point, simple dimension-counting tells us we need $S\, \gsim \, 4^N$,  so that we have enough `fine motor skills' to fill out the dimensionality of the neighbourhood of a point. But even for $S = 4^N$, if $\delta$ is too small we don't have enough `gross motor skills' to reach every $\epsilon$-ball. Consider starting with a complexity geometry path, and `coarse-graining' into a zig-zag$_\delta$ path by divvying up the path into segments of length $\delta$ and within each segment applying the time-average Hamiltonian. This introduces a per-segment inner-product error of about $\delta^2$, for a total inner-product error of $S \delta^2$. This suggests that the critical values are $S = 4^N$ and $\delta = 2^{-N}$, and that the zig-zag$_\delta$ diameter of the unitary group is $4^N \delta$ for $\delta >2^{-N}$ and $2^N$ for $\delta < 2^{-N}$. Taking $\delta \rightarrow 0$ recovers Eq.~\ref{eq:conjecture}. \\


 Nielsen's original vision was to use the tools of differential geometry to lowerbound the complexity of the complexity geometry, and then use that to prove novel lowerbounds on gate complexity. This paper has realized the first half of that vision. When we proved complexity lowerbounds for the complexity geometry, we also via Eq.~\ref{eq:geomlessthangatesobvs}  implicitly proved lowerbounds for gate complexity.  But those lowerbounds were too weak to be novel. The problem is that the gate definition of complexity is so much less permissive than the complexity geometry definition that it is easier to directly prove limitations on gate complexity rather than to indirectly prove limitations via the complexity geometry. The Bishop-Gromov bound has proven itself a powerful tool for establishing complexity lowerbounds for the complexity geometry, but it seems we're going to need more powerful tools still to realize the totality of Nielsen's vision.

\section*{Acknowledgments }

I'd like to thank Henry Lin, Leonard Susskind, and particularly Michael Freedman, as well as wikipedia for alerting me to the existence of the Bishop-Gromov bound.

\appendix

\section{The curvature of right-invariant metrics}  \sc
Milnor \cite{Milnor} gives as his Lemma 1.1 that, defining $\alpha_{IJK} \equiv 
 \mathcal{I}_k \textrm{Tr}[ [\sigma_I,\sigma_J] \sigma_K]$, the sectional curvature  is  
\begin{equation} \label{eq:milnorsformulation}
\kappa (\sigma_I, \sigma_J) = \frac{1}{4} \sum_{\sigma_K} 2 \alpha_{IJK} \left( - \alpha_{IJK} + \alpha_{JKI} + \alpha_{KIJ}  \right)  - (\alpha_{IJK} - \alpha_{JKI} + \alpha_{KIJ}) (\alpha_{IJK} + \alpha_{JKI}  - \alpha_{KIJ}) 
\end{equation}
plus a term proportional to $\alpha_{KII}\alpha_{KJJ}$ that vanishes for U($2^N$) because the structure constants 
are completely antisymmetric. Indeed, using the antisymmetry, as a matter of algebra this is 
\begin{equation}
\kappa (\sigma_I, \sigma_J) =  
  \sum_{\sigma_K} \frac{\textrm{Tr}[[\sigma_I,\sigma_J ]  \sigma_K]^2}{4\mathcal{I}_{\sigma_I} \mathcal{I}_{\sigma_J} \mathcal{I}_{\sigma_K}}  \left(  - 3\mathcal{I}_{\sigma_K}^2 + 2 \mathcal{I}_{\sigma_K}  (\mathcal{I}_{\sigma_I}  + \mathcal{I}_{\sigma_J} )  + (\mathcal{I}_{\sigma_I} - \mathcal{I}_{\sigma_J})^2  \right) \ . \label{eq:sectionsinniceform}
\end{equation}
The Ricci curvature is diagonal in the generalized Pauli basis, as proved in Appendix A.3.~of Ref.~\cite{Nielsen4}. Along the diagonal, the Ricci curvature in a direction is the sum of the sectional curvatures of all sections with a leg pointing down that direction
\begin{equation}
\mathcal{R}_{\sigma_I}^{\ \sigma_I} \Bigl|_\textrm{not summed}= \sum_{\sigma_J} \mathcal{R}_{\sigma_I \sigma_J}^{\ \ \ \  \sigma_J \sigma_I} \Bigl|_\textrm{not summed over $\sigma_I$}  = \sum_{\sigma_J} \kappa (\sigma_I, \sigma_J). \label{eq:definitionofricciintermsofsectionals}
\end{equation}
Plugging Eq.~\ref{eq:sectionsinniceform} into \ref{eq:definitionofricciintermsofsectionals}, and then symmetrizing over $\sigma_J$ \& $\sigma_K$ (allowed since they are both inside the sum), and then  using that $\textrm{Tr}[[\sigma_I,\sigma_J ]  \sigma_K]= \textrm{Tr}[[\sigma_I,\sigma_J ]  \sigma_K]  \delta_{(\sigma_K)( [\sigma_I,\sigma_J ])}$, gives Eq.~\ref{eq:tobereferredtolater}.

\end{document}